# Overcoming the time limitation in Molecular Dynamics simulation of crystal nucleation: a persistent-embryo approach

Yang Sun[1*], Huajing Song[1], Feng Zhang[1†], Lin Yang[1], Zhuo Ye[1], Mikhail I. Mendelev[1], Cai-Zhuang Wang[1,2], Kai-Ming Ho[1,2,3]

[1]Ames Laboratory, US Department of Energy, Ames, Iowa 50011, USA

[2]Department of Physics, Iowa State University, Ames, Iowa 50011, USA

[3]Hefei National Laboratory for Physical Sciences at the Microscale and Department of Physics, University of Science and Technology of China, Hefei, Anhui 230026, China

(Dated: February 23, 2018)

Abstract

The crystal nucleation from liquid in most cases is too rare to be accessed within the limited timescales of the conventional molecular dynamics (MD) simulation. Here, we developed a "persistent embryo" method to facilitate crystal nucleation in MD simulations by preventing small crystal embryos from melting using external spring forces. We applied this method to the pure Ni case for a moderate undercooling where no nucleation can be observed in the conventional MD simulation, and obtained nucleation rate in good agreement with the experimental data. Moreover, the method is applied to simulate an even more sluggish event: the nucleation of the B2 phase in a strong glass-forming Cu-Zr alloy. The nucleation rate was found to be 8 orders of magnitude smaller than Ni at the same undercooling, which well explains the good glass formability of the alloy. Thus, our work opens a new avenue to study solidification under realistic experimental conditions via atomistic computer simulation.

[*]yangsun@ameslab.gov
[†]fzhang@ameslab.gov

Homogeneous crystal nucleation from an undercooled liquid is a fundamental process that plays an important role in numerous areas ranging from materials science to biophysics [1]. In the classical nucleation theory (CNT), the nucleation is described as a competition between the energy gain associated with the transformation of the bulk liquid into a crystal phase and the energy cost of creating a solid-liquid interface such that the change in the free energy associated with the formation of a nucleus containing $N$ atoms can be presented as:

$$\Delta G(N) = N\Delta\mu + s(N/\rho)^{2/3}\gamma, \quad (1)$$

where $\rho$ is the atomic density, $\Delta\mu$ ($< 0$) is the chemical potential difference between the bulk solid and liquid, $\gamma$ ($> 0$) is the solid-liquid interfacial free energy and $s$ is a factor to account for the nucleus shape. As schematically shown in Fig. 1(a), this competition between the bulk and interface terms leads to a critical barrier $\Delta G^*$ where the nucleus reaches the critical size $N^*$. The low probability of overcoming this free energy barrier makes it inefficient to sample nucleation events in conventional MD simulations [2]. To circumvent this difficulty, advanced sampling techniques such as umbrella sampling [3–5] and metadynamics [6] can be used. With the help of biased potentials, these techniques can in principle map out the free energy barrier for nucleation. However, they do not directly give the correct kinetics of the unbiased system; and thus other methods, such as kinetic Monte Carlo (KMC), have to be used to obtain necessary kinetic parameters for evaluating the nucleation rate [7], which significantly adds to the complexity of the problem. The critical nucleus size was also determined by embedding a large crystal cluster into the liquid and watching if the cluster grows or disappears [8]. Although this method can provide a fast estimation of the critical nucleus size [9–11], the initial equilibration process during which the cluster should melt, can lead to a considerable overestimation of the critical nucleus size [12]. Moreover, the artificially chosen initial cluster can lead to an unreal description of the nucleus shape, such as the non-spherical nucleus shape in the Lennard-Jones system reviewed recently by Sosso *et al.* in Ref. [2].

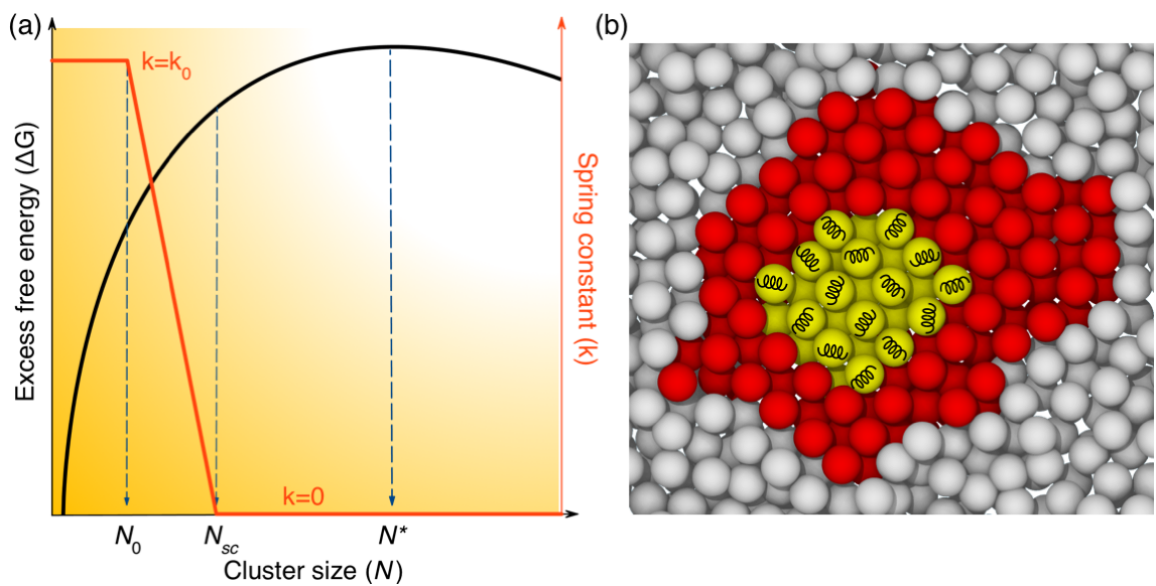

FIG. 1. Using the persistent-embryo method to reach the critical nucleus. (a) The excess free energy (black) and spring constant (red) as a function of the crystalline cluster size $N$. $N_0$ is the number of atoms in the constrained embryo. The red curve shows that the strength of the spring constant decreases with the increasing cluster size. The spring is completely removed when the cluster size reaches the threshold value $N_{sc}$. $N^*$ is the critical size. (b) A cross-section of the as-grown crystalline cluster around embryo. The yellow atoms with spring icon are the persistent embryo. The red are the as-grown atoms, showing the crystalline packing. The gray are the liquid atoms.

In the present study, we took advantage of the well-known fact that the dependence of $\Delta G$ on $N$ has a convex shape (see Fig. 1a), which means that a large fraction of $\Delta G^*$ must be overcome only to grow a small crystalline cluster (embryo). Thus, if the embryo can be kept from re-melting, it can reach the critical size even during a relatively short MD simulation. Therefore, we propose a persistent-embryo method to achieve this, in which external spring forces are applied to constrain the embryo from melting. First, we create a crystalline embryo with $N_0$ atoms ($N_0$ is much smaller than $N^*$), which is then inserted into the liquid while a *tunable* harmonic potential is added to each atom in the embryo to effectively keep it from melting. As the embryo grows, the harmonic potential is gradually weakened and is completely removed when the cluster size reaches a sub-critical threshold $N_{sc}$ ($< N^*$): the spring constant corresponding to the harmonic potential is set as $k(N) = k_0 \frac{N_{sc}-N}{N_{sc}}$ if $N < N_{sc}$ and $k(N) = 0$ otherwise. If the nucleus melts the harmonic potential is gradually enforced. The strategy to adjust the spring constant to zero before reaching the critical nucleus size ensures the dynamics of the system is unbiased at the critical point, which is an advantage of this approach compared to others such as the lattice mold method [13]. A schematic of the simulation configuration is shown in Fig. 1(b). We emphasize since the springs are removed well before the nucleus reaches the critical size, the overall process simulates homogeneous nucleation.

During the MD simulation, the *NPT* ensemble is applied with Nose-Hoover thermostats. The time step of the simulation is 1.0 *fs*. The sample size is set up to 32,000 atoms which is at least 10 times larger than the critical nucleus size. The Finnis-Sinclair (FS) potentials [14] were used for the investigation of Ni [15] and CuZr [16] systems. These FS potentials were developed to accurately reproduce the melting point data and the liquid structure. The initial liquid is equilibrated for 1 *ns*. The embryo is inserted in the liquid by removing liquid atoms that are closer to the embryo atoms than 2.0 Å. All the simulations were performed using the GPU-accelerated LAMMPS code [17–19]. To quickly identify the solid-like and liquid-like atoms during MD simulation, the widely-used bond-orientational order parameter [20,21] is employed by calculating $S_{ij} = \sum_{m=-6}^{6} q_{6m}(i) \cdot q_{6m}^*(i)$ between two neighboring atoms based on the Steinhardt parameter $q_{6m}(i) = \frac{1}{N_b(i)} \sum_{j=1}^{N_b(i)} Y_{lm}(\vec{r}_{ij})$, where $Y_{lm}(\vec{r}_{ij})$ is the spherical harmonics and $N_b(i)$ is the number of nearest neighbors of atom $i$. Two neighboring atoms $i$ and $j$ are considered to be connected when $S_{ij}$ exceeds a threshold. The threshold is carefully chosen based on Espinosa *et al.*'s "equal mislabeling" method [11], which gives the lowest probability to mislabel the liquid and solid (see

Supplemental Material [12] for details). The atoms with 6 connected neighbors are recognized as solid-like. Then the cluster analysis [22], which uses the crystalline bond length as the cutoff distance to choose neighbor atoms, is applied to measure the size of the solid cluster which formed around the initial embryo.

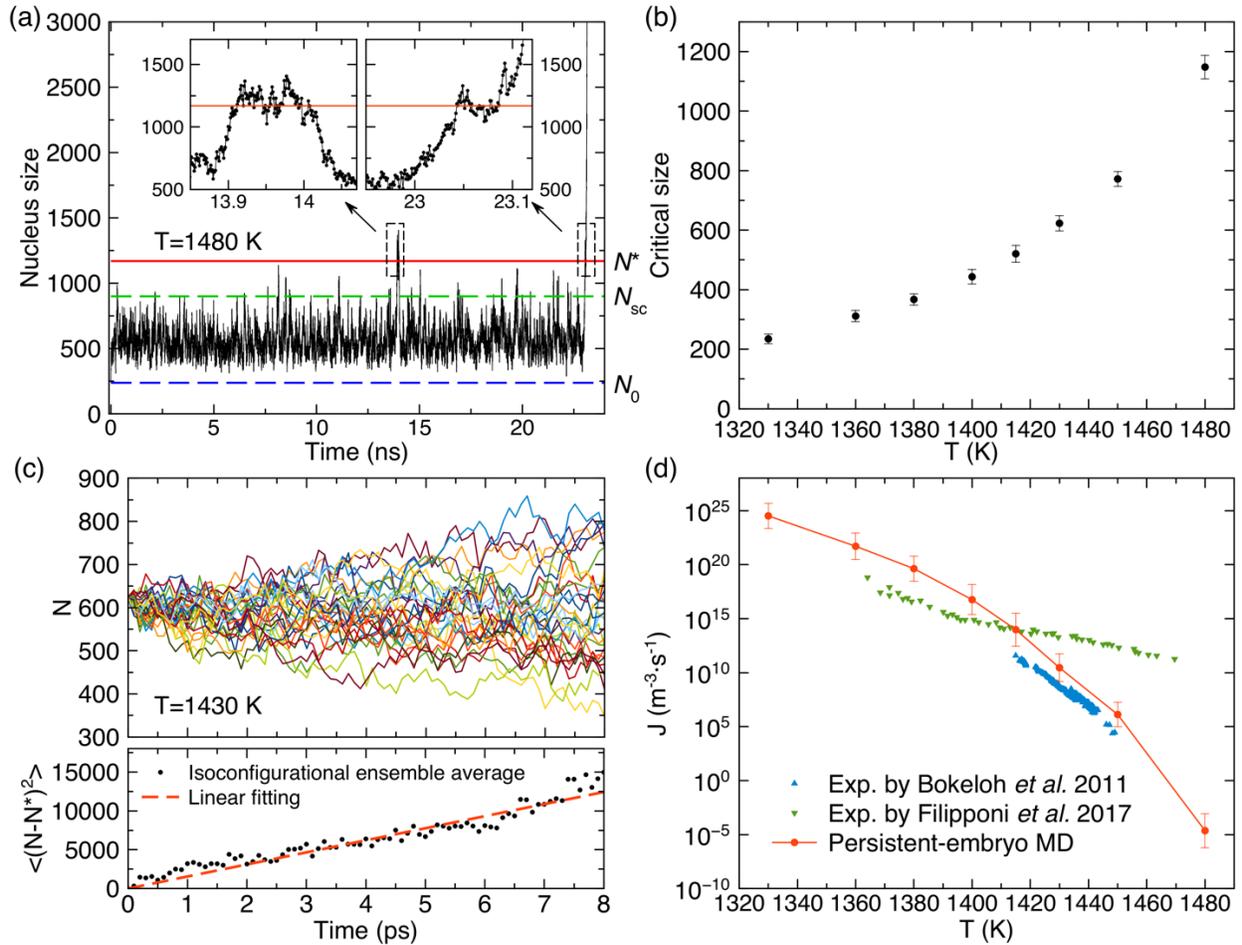

FIG. 2. The persistent-embryo MD simulation of the crystal nucleation in the undercooled liquid Ni. (a) The nucleus size versus time of one Ni nucleation trajectory at 1480 K. The blue dashed line shows the atom number $N_0$ in the persistent embryo. The green dashed line indicates the threshold to remove the spring and the red solid line indicates the critical size $N^*$. Two inserts zoom in two plateaus at the critical size. (b) The critical size as a function of the temperature. (c) The upper panel shows the nucleus size versus time for the isoconfigurational ensemble with 30 MD runs. Each color indicates an independent MD trajectory. The bottom panel shows the ensemble average of $|\Delta N^*(t)|^2 = |N(t) - N^*|^2$. The dashed line indicates the linear fitting to the first 5 ps to derive the attachment rate; (d) The nucleation rate as a function of the temperature for Ni. The simulation results are connected to guide the eye. The experimental data are from Ref. [23] (▲) and Ref. [24] (▼).

We first applied this method to the pure Ni case with a wide range of moderate undercooling. Under these conditions, experimental nucleation events occur on the time scale of seconds [23] and, hence, cannot be observed in the conventional MD simulation with the simulation time usually less than 1 microsecond. With the help of a persistent embryo, the long-time fluctuation of the nucleation with nucleus smaller than the embryo is suppressed. The barrier to be overcome by the simulation is reduced so that the nucleation can be observed at the typical MD timescale. When the nucleus reaches the critical size, it has equal chance to dissolve or further grow. Thus, one should expect that the size of the nucleus will fluctuate about $N^*$ within an extended time, which will result in a plateau at the critical region on the $N(t)$ curve. This unique signal can help us to accurately measure $N^*$ in our simulations. We, therefore, launched multiple independent MD runs (up to 50 runs) to collect such critical plateaus for statistical analysis. An example is shown in Fig. 2(a) (see more examples in Supplemental Material [12]). Although the length of the plateaus varies in different runs, their heights are almost identical. Thus, the critical size can be determined statistically by averaging over all the plateau heights. The obtained critical nucleus size as a function of temperature is shown in Fig. 2(b). We note that as long as the $N_0$ and $N_{sc}$ are chosen such that the fluctuating plateau can be observed within the typical MD timescale in the simulations, different choices of the embryo shape, $N_0$ and $N_{sc}$ give a consistent measurement of the critical nucleus size (see Supplemental Material for details [12]). The fundamental reason that the persistent-embryo method allows an accurate measurement of the critical nucleus size is that one can observe the actual fluctuations of a critical nucleus in an unbiased environment, and perform extensive statistical analysis based on these fluctuations. This unique feature will be even more important for treating stoichiometric compounds with larger anisotropy of the interfacial properties [25].

In the CNT, the nucleation rate, $J$, can be expressed as $J = \kappa \exp(-\Delta G^*/k_B T)$, where $k_B$ is the Boltzmann constant, and $\kappa$ is a kinetic prefactor. $\Delta G^*$ is related to the driving force $|\Delta\mu|$ and the critical size $N^*$ as $\Delta G^* = \frac{1}{2}|\Delta\mu|N^*$ (see Supplemental Material [12]). Using the steady-state model to derive the kinetic prefactor [1], we can express the nucleation rate as

$$J = \rho_L f^+ \sqrt{\frac{|\Delta\mu|}{6\pi k_B T N^*}} \exp\left(-\frac{|\Delta\mu|N^*}{2k_B T}\right), \quad (2)$$

where $f^+$ is the attachment rate of a single atom to the critical nucleus and $\rho_L$ is the liquid density. $\Delta\mu$ can be computed by integrating the Gibbs-Helmholtz equation from the undercooling temperature to the melting point [26]. Following the pioneering work by Auer and Frenkel [7], once the critical nucleus is available, the attachment rate can be measured with MD simulation as the effective diffusion constant for the change in critical nucleus size: $f^+ = \frac{\langle|\Delta N^*(t)|^2\rangle}{2t}$. Figure 2(c) shows the measurement of the attachment rate at the critical nucleus using an isoconfigurational ensemble [27]. 30 independent MD runs were performed starting from the same atomic configuration with a critical nucleus but with atomic momenta randomly assigned using the Maxwell distribution. As there are no constraints in the embryos anymore, the critical nucleus indeed melted in half of the MD runs and grew in the other half runs, which further validates the determination of the critical nucleus size.

Figure 2(d) shows that the nucleation rate in pure Ni as a function of temperature. The nucleation rate computed with the persistent-embryo MD covers a wide undercooling range, which can be compared directly to the recent experimental measurements [23,24]. The results agree well with Bokeloh *et al.*'s experimental measurements from 1400 K to 1450 K, in which homogeneous nucleation was carefully probed [23]. Our results slightly deviate from Filipponi *et al.*'s measurements [24] from 1360 K to 1380 K but these data could be affected by possible heterogeneous nucleation [24].

Compared to the pure Ni case, it is a much more challenging task to simulate a nucleation in a glass-forming alloy, because the crystal nucleation can be bypassed even on the experimental timescale in such a system. Here, we employ the persistent-embryo method to simulate the B2 phase nucleation in the $Cu_{50}Zr_{50}$ alloy, which has attracted extensively attention as a strong binary glass former [28,29]. As shown in Fig. 3(a), we can still obtain the critical nucleus size by sampling plateaus on $N(t)$ curves collected in different MD runs. It is interesting to note that the plateau can sustain much longer time in CuZr than in Ni. This can be attributed to a much slower attachment/detachment rate, which was measured in isoconfigurational simulations shown in Fig. 3 (b).

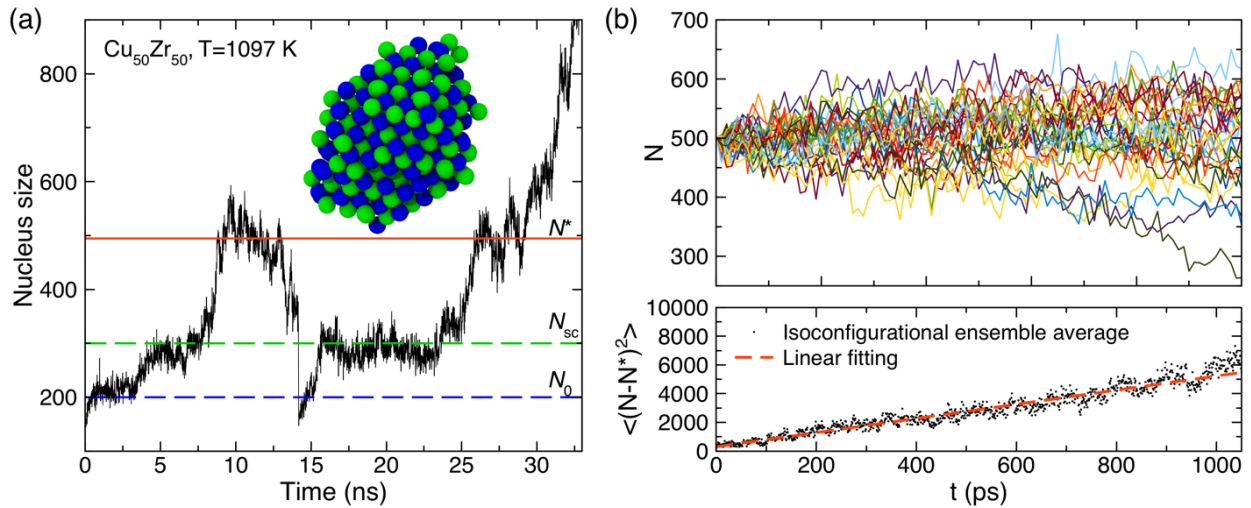

FIG. 3. The persistent-embryo MD for B2 nucleation in $Cu_{50}Zr_{50}$ undercooled liquid at 1097 K. (a) Nucleus size as a function of time. The insert shows the B2 critical nucleus. (b) 30 MD runs starting from the configuration with critical nucleus are performed. Each color indicates an independent MD trajectory. The dashed line shows the linear fitting of the ensemble average to derive the attachment rate.

The nucleation rate of the B2 phase from the $Cu_{50}Zr_{50}$ liquid alloy was found to be 8 orders of magnitude smaller than the nucleation of the FCC phase in liquid Ni. This explains why the $Cu_{50}Zr_{50}$ liquid can bypass the crystal phase and be driven to the glassy state when cooled at a sufficient fast rate. In Table.1, we compare several factors that could affect the nucleation rates in liquid Ni and $Cu_{50}Zr_{50}$ alloy at the same undercooling $T' = (T_m - T)/T_m$. The examination of this table shows that 6 orders of magnitude are caused by the higher free energy barrier and remaining 2 orders of magnitude result from the smaller attachment rate. The higher nucleation barrier of the B2 phase in the $Cu_{50}Zr_{50}$ alloy is associated with much larger energy penalty of forming the liquid/B2 interface comparing to that for the liquid/fcc interface in Ni (see Supplemental Material for details [12]). Note the diffusivities of Ni and $Cu_{50}Zr_{50}$ are quite similar. Thus, the attachment rate may be highly affected by the structure of solid-liquid interface as observed by Tang and Harrowell [30].

Table 1 The critical nucleus size ($N^*$), free energy barrier contribution ($e^{-\Delta G^*/k_B T}$), attachment rate ($f^+$), prefactor ($\kappa$), nucleation rate ($J$) and atomic diffusivity ($D$) for the pure Ni and $Cu_{50}Zr_{50}$ liquid alloy at same undercooling $T' = \frac{T_m - T}{T_m}$.

| System | T (K) | $T'$ | $N^*$ | $e^{-\Delta G^*/k_B T}$ | $f^+$ ($s^{-1}$) | $\kappa$ ($m^{-3} s^{-1}$) | $J$ ($m^{-3} s^{-1}$) | $D$ ($m^2/s$) |
|---|---|---|---|---|---|---|---|---|
| Ni | 1430 | 17% | 623 | $1.0 \times 10^{-31}$ | $7.6 \times 10^{14}$ | $2.8 \times 10^{41}$ | $2.9 \times 10^{10}$ | $2.0 \times 10^{-9}$ |
| $Cu_{50}Zr_{50}$ | 1097 | 17% | 495 | $9.8 \times 10^{-38}$ | $2.5 \times 10^{12}$ | $8.3 \times 10^{38}$ | 81.3 | Cu: $9.8 \times 10^{-10}$<br>Zr: $7.2 \times 10^{-10}$ |

In summary, the proposed persistent-embryo method dramatically extends the ability of the MD simulation to explore the rare nucleation without the use of biasing forces near the critical point. The spontaneously formed critical nucleus, the critical size and the kinetic prefactor can be measured so that the nucleation rate can be computed in the CNT framework. The study of the nucleation in pure Ni demonstrated a good agreement with available experimental data proofing the reliability of the preformed work. The investigation of the nucleation in the $Cu_{50}Zr_{50}$ liquid alloy revealed an extremely low nucleation rate which explains the high glass formability of this alloy. These successes demonstrate that our work opens a practical way to quantitatively estimate nucleation rates under realistic experimental conditions.

**Reference**

[1]  K. F. Kelton and A. L. Greer, *Nucleation in Condensed Matter: Application in Materials and Biology* (Elsevier, Amsterdam, 2010).

[2]  G. C. Sosso, J. Chen, S. J. Cox, M. Fitzner, P. Pedevilla, A. Zen, and A. Michaelides, Chem. Rev. **116**, 7078 (2016).

[3]  G. M. G. M. Torrie and J. P. J. P. Valleau, J. Comput. Phys. **23**, 187 (1977).

[4]  S. Kumar, J. M. Rosenberg, D. Bouzida, R. H. Swendsen, and P. A. Kollman, J. Comput. Chem. **13**, 1011 (1992).

[5]  S. Auer and D. Frenkel, Nature **409**, 1020 (2001).

[6]  A. Laio and M. Parrinello, Proc. Natl. Acad. Sci. U. S. A. **99**, 12562 (2002).

[7]  S. Auer and D. Frenkel, J. Chem. Phys. **120**, 3015 (2004).

[8]  X.-M. Bai and M. Li, J. Chem. Phys. **122**, 224510 (2005).

[9]  E. Sanz, C. Vega, J. R. Espinosa, R. Caballero-Bernal, J. L. F. Abascal, and C. Valeriani, J. Am. Chem. Soc. **135**, 15008 (2013).

[10] T. Mandal and R. G. Larson, J. Chem. Phys. **146**, 134501 (2017).

[11] J. R. Espinosa, C. Vega, C. Valeriani, and E. Sanz, J. Chem. Phys. **144**, 034501 (2016).

[12] For the Ni potential studied in the present paper, the conventional seeding method systematically overestimates the critical nucleus size which leads to an error of ~3 orders of magnitudes on the nucleation rate compared to the persistent-embryo method and the burte-force MD. See Supplemental Material [url] for the comparison between the seeding and current methods, additional critical plateaus and computational details, which includes Refs. [31-33].

[13] J. R. Espinosa, P. Sampedro, C. Valeriani, C. Vega, and E. Sanz, Faraday Discuss. **195**, 569 (2016).

[14] M. W. Finnis and J. E. Sinclair, Philos. Mag. A **50**, 45 (1984).

[15] M. I. Mendelev, M. J. Kramer, S. G. Hao, K. M. Ho, and C. Z. Wang, Philos. Mag. **92**, 4454 (2012).

[16] M. I. Mendelev, M. J. Kramer, R. T. Ott, D. J. Sordelet, D. Yagodin, and P. Popel, Philos. Mag. **89**, 967 (2009).

[17] W. M. Brown, P. Wang, S. J. Plimpton, and A. N. Tharrington, Comput. Phys. Commun. **182**, 898 (2011).

[18] W. M. Brown, A. Kohlmeyer, S. J. Plimpton, and A. N. Tharrington, Comput. Phys. Commun. **183**, 449 (2012).

[19] W. M. Brown and M. Yamada, Comput. Phys. Commun. **184**, 2785 (2013).

[20] P. J. Steinhardt, D. R. Nelson, and M. Ronchetti, Phys. Rev. B **28**, 784 (1983).

[21] P. Rein ten Wolde, M. J. Ruiz-Montero, and D. Frenkel, J. Chem. Phys. **104**, 9932 (1996).

[22] P.-N. Tan, M. Steinbach, and V. Kumar, *Introduction to Data Mining* (Pearson Addison Wesley, 2005).

[23] J. Bokeloh, R. E. Rozas, J. Horbach, and G. Wilde, Phys. Rev. Lett. **107**, 145701 (2011).

[24] A. Filipponi, A. Di Cicco, S. De Panfilis, P. Giammatteo, and F. Iesari, Acta Mater. **124**, 261 (2017).

[25] S. R. Wilson and M. I. Mendelev, Philos. Mag. **95**, 224 (2015).

[26] M. I. Mendelev, M. J. Kramer, C. A. Becker, and M. Asta, Philos. Mag. **88**, 1723 (2008).

[27] A. Widmer-Cooper, P. Harrowell, and H. Fynewever, Phys. Rev. Lett. **93**, 135701 (2004).

[28] W. H. Wang, J. J. Lewandowski, and A. L. Greer, J. Mater. Res. **20**, 2307 (2005).

[29] Y. Li, Q. Guo, J. A. Kalb, and C. V. Thompson, Science **322**, 1816 (2008).

[30] C. Tang and P. Harrowell, Nat. Mater. **12**, 507 (2013).

[31] F. C. Frank, Proc. R. Soc. A **215**, 43 (1952).

[32] J. R. Espinosa, C. Vega, C. Valeriani, and E. Sanz, J. Chem. Phys. **142**, 194709 (2015).

[33] G. Chkonia, J. Wölk, R. Strey, J. Wedekind, and D. Reguera, J. Chem. Phys. **130**, 64505 (2009).

**Acknowledgements**

We thank M. J. Kramer, R. E. Napolitano, X. Song and R. T. Ott from Ames Laboratory for valuable discussion. Work at Ames Laboratory was supported by the US Department of Energy, Basic Energy Sciences, Materials Science and Engineering Division, under Contract No. DE-AC02-07CH11358, including a grant of computer time at the National Energy Research Supercomputing Center (NERSC) in Berkeley, CA. K.M.H. acknowledges support from USTC Qian-Ren B (1000-Talents Program B) fund. The Laboratory Directed Research and Development (LDRD) program of Ames Laboratory supported the use of GPU computing.